\documentclass[reqno,12pt]{amsart}

\usepackage[centertags]{amsmath}
\usepackage{amsfonts}
\usepackage{amssymb}
\usepackage{amsthm}
\usepackage{newlfont}

\theoremstyle{plain}
  \newtheorem{theorem}{Theorem}[section]
  \newtheorem{corollary}[theorem]{Corollary}
  \newtheorem{proposition}[theorem]{Proposition}
  \newtheorem{lemma}[theorem]{Lemma}
\theoremstyle{definition}
  \newtheorem{definition}{theorem}[section]
  
\theoremstyle{remark}
  \newtheorem{remark}[theorem]{Remark}

\numberwithin{equation}{section}

\newcommand{\opunit}{\text{1}\kern-0.22em\text{l}}

\newcommand{\bsZ}{{\boldsymbol Z}}

\newcommand{\be}{\begin{equation}}
\newcommand{\ee}{\end{equation}}

\newcommand{\bl}{\begin{lemma}}
\newcommand{\el}{\end{lemma}}

\newcommand{\br}{\begin{remark}}
\newcommand{\er}{\end{remark}}

\newcommand{\bt}{\begin{theorem}}
\newcommand{\et}{\end{theorem}}

\newcommand{\bd}{\begin{definition}}
\newcommand{\ed}{\end{definition}}

\newcommand{\bp}{\begin{proposition}}
\newcommand{\ep}{\end{proposition}}

\newcommand{\bc}{\begin{corollary}}
\newcommand{\ec}{\end{corollary}}

\newcommand{\bpr}{\begin{proof}}
\newcommand{\epr}{\end{proof}}

\newcommand{\bi}{\begin{itemize}}
\newcommand{\ei}{\end{itemize}}

\newcommand{\ben}{\begin{enumerate}}
\newcommand{\een}{\end{enumerate}}

\newcommand{\R}{\mathbb R}
\newcommand{\N}{\mathbb N}

\newcommand{\pee}{\mathbb P}

\newcommand{\re}{\ensuremath{\mathcal{R}}}

\newcommand{\vi}{\ensuremath{\varphi}}

\newcommand{\si}{\ensuremath{\sigma}}

\newcommand{\epsi}{\ensuremath{\epsilon}}

\def\now{
\ifnum\time<60
          12:\ifnum\time<10 0\fi\number\time am
          \else
            \ifnum\time>719\chardef\a=`p\else\chardef\a=`a\fi
          \hour=\time
          \minute=\time
          \divide\hour by 60 
          \ifnum\hour>12\advance\hour by -12\advance\minute by-720 \fi
          \number\hour:%
          \multiply\hour by 60 
          \advance\minute by -\hour
          \ifnum\minute<10 0\fi\number\minute\a m\fi}
\newcount\hour
\newcount\minute
\numberwithin{equation}{section}         

\theoremstyle{remark}

\begin{document}

\begin{center}
\noindent{\large \bf The Potts model built on sand} \\

\vspace{15pt}

{\bf E. Dinaburg}\footnote{Institute of Physics of Earth, RAS,
Moscow}, {\bf C. Maes}\footnote{Instituut voor Theoretische
Fysica, K.U.Leuven\\ email: christian.maes@fys.kuleuven.ac.be},
{\bf S. Pirogov}\footnote{Institute for Information Transmission
Problems, RAS, Moscow, supported by RFFI-grant 02-01-01276}, {\bf
F. Redig}\footnote{Faculteit Wiskunde en Informatica, Technische
Universiteit Eindhoven}, {\bf A. Rybko}\footnote{Institute for
Information Transmission Problems, RAS, Moscow, supported by
RFFI-grant 02-01-00068}

\end{center}
\vspace{20pt} \footnotesize \noindent {\bf Abstract}: We consider
the $q=4$ Potts model on the square lattice with an additional
hard-core nonlocal interaction.  That interaction arises from the
choice of the reference measure taken to be the uniform measure on
the recurrent configurations for the abelian sandpile model. In
that reference measure some correlation functions have a power-law
decay.
 We investigate the low-temperature phase
diagram and we prove the existence of a single stable phase
 with exponential decay of correlations.  For all boundary
 conditions
the density of 4 in the infinite volume limit goes to one as the
temperature tends to zero.

 \normalsize \vspace{12pt}
\renewcommand{\baselinestretch}{2}\normalsize

\section{Model}
We define the model on the two-dimensional lattice $\bsZ^2$.
Lattice sites are denoted by $x,y,z$ and we write $x\sim y$ if $x$
and $y$ are nearest neighbors. For a subset $V\subset\bsZ^2$ we
denote by $\partial V$ the exterior boundary of $V$, i.e., the set
of those $y\in\bsZ^2\setminus V$ such that there exists a nearest
neighbor of $y$ in $V$, $\overline{V}=V\cup\partial V$, and the
set $\partial(\bsZ^2\setminus V)$ is called the inner boundary of
$V$. The square $[-n,n]^2\cap\bsZ^2$ is denoted by $V_n$. The
configuration space is $\Omega = \{ 1,2,3,4 \}^{\bsZ^2}$. Elements
of $\Omega$ are written as $\eta,\xi$. For a configuration $\eta$,
$\eta(x) \in \{ 1,2,3,4\}$ is interpreted as the ``number of sand
grains" at $x$. For $n\in\N$, $\Omega_n= \{1,2,3,4\}^{V_n}$
denotes the set of finite volume height configurations. Below we
introduce measures $\mu^a_{\beta, n}$ on $\Omega_n$ that
correspond to the finite volume Potts model at inverse temperature
$\beta$ with boundary condition $a\in\{1,2,3,4\}$, restricted to a
special set of ``recurrent configurations" defined from the
abelian sandpile model (cf. subsection \ref{sand} below).  Our
main result is that for $\beta$ large, $\mu^4_{\beta,n}$ forms the
single stable phase of that model.

\subsection{Potts model}
The Potts Hamiltonian with fixed boundary condition $a\in
\{1,2,3,4\}$ on the volume $V_n$ is
\[
H_n(\eta|a) = \sum_{x\sim y \in V_n}  I[\eta(x) \neq
\eta(y)]
\]
That is a finite sum over nearest neighbor pairs of sites of which at least one
belongs to $V_n$ and where it is understood that we substitute
$\eta(z)=a$ whenever $z\notin V_n$.

The Hamiltonians $H_n(\eta|a)$ give rise to the finite volume Gibbs measures on $\Omega_n$:
\[
\nu_{\beta,n}^a(\eta) = \frac 1{ Z_{\beta,n}^a}\, \, \exp[-\beta
H_n(\eta|a)]
\]
where $\beta>0$ is
 the inverse temperature and where the normalizing factor $Z_{\beta,n}^a$ is
the partition function.  It is well-known that there exists a
critical inverse temperature $\beta_c \in (0,+\infty)$ such that
for $\beta < \beta_c$ the Potts model has a unique infinite volume
Gibbs measure (as $n\uparrow +\infty$), i.e., the weak limits of
$\nu_{\beta,n}^a$ for different $a$ coincide, while for $\beta <
\beta_c$, these weak limits are all different and define 4
mutually singular ergodic Gibbs measures $\nu_\beta^a$ on
$\Omega$, called the pure phases.

\subsection{Sandpile model}\label{sand}
The abelian sandpile model in volume $V_n$ is a Markov  chain on
$\Omega_n$. We briefly introduce that Markov chain, more details
can be found in the original paper \cite{btw}, and in \cite{dhar} or \cite{meest}.\\
The Markov chain starting from $\eta_0\in\Omega_n$ is defined as
follows. Suppose that $\eta_{t-1}$ is the configuration at time
$t-1\geq 0$. Pick randomly a site in $V_n$, say $x\in V_n$, and
add one grain at $x$ to $\eta_{t-1}$.  In case $\eta_{t-1}(x) \leq
3$, the new configuration is simply
\[
\eta_t(y) = \eta_{t-1}(y) + \delta_{x,y}
\]
with $\delta_{x,y}$ the Kronecker delta.  In case $\eta_{t-1}(x) = 4$, by adding one grain
at $x$ the number of grains at $x$ becomes equal to 5.  That site will now topple, i.e.,
$4$ grains are removed from $x$ and one grain is given
to each neighbor of $x$ in $V_n$. At the boundary, grains are
lost when the site topples.  It is now possible that the number of grains at
one or more neighbors of $x$ exceeds 4 and we have to repeat the toppling
operation on all of these, and so on.  It turns out that no matter in what
order we perform these toppling operations, at the end of the avalanche
a unique configuration $\eta_t \in \Omega_n$ appears.  In that way, a
discrete time Markov chain on $\Omega_n$ is defined where the only
randomness is in the independently repeated uniform choice of the site
where a grain is added.

Analysis of that Markov chain learns that it has a unique class
$\re_n$ of recurrent configurations and the stationary measure $\lambda_n$
is uniform
on that class:
\[
\lambda_n(\eta) = \frac 1{|\re_n|} I[\eta \in \re_n]
\]
see \cite{dhar}.

Whether a particular configuration $\eta\in\Omega_n$ belongs to
$\re_n$ can be decided from the output of the so-called burning
algorithm \cite{dhar}. The burning algorithm has as an input the
configuration $\eta$ and its output is a set $A\subset V_n$. It
runs as follows: start from $A_0= V_n$ and remove (``burn") all
those vertices $x\in A_0$ (and edges containing $x$) which satisfy
$\eta(x) > n_{A_0} (x)$ where $n_V (x)$ denotes the number of
neighbors of $x$ in $V$. This gives $A_1$; now proceed in the same
way with $A_1$, etc. until no further vertices can be removed. The
output $A$ of the algorithm is the set of remaining vertices.
Recurrence is then characterized by ``burnability", i.e.,
$\eta\in\re_n$ if and only if $A=\emptyset$, i.e., all vertices
can be burned.

The stationary measure $\lambda_n$ is thus the uniform probability
measure on all burnable configurations in $V_n$. The cardinality
$|\re_V|$ (= the number of recurrent configurations in $V$) equals
the determinant of the discrete Laplacian on $V$ with open
boundary conditions see \cite{dhar}.  E.g. if $V$ is a square
containing $N$ sites, then $|\re_V| \simeq (3.21)^{N}$.

For a proof of these facts, see e.g. \cite{dhar}, \cite{IP},
\cite{meest} or \cite{Speer}.
 Remark that for all finite $V\subset\bsZ^2$,
the constant configurations $\eta\equiv 4$ and
$\eta\equiv 3$ are in $\re_V$, but
$\eta\equiv 2$ and $\eta\equiv 1$ are not recurrent except for some very special
 choices of $V$. One easily concludes that the condition
 that $\eta\in\re_n$ is a nonlocal hard-core
constraint.

The following proposition is an immediate consequence of the
burning algorithm.

\begin{proposition}\label{domi}
If $\eta \in \re_n$ and $\xi \geq \eta$ (pointwise), then  $\xi
\in \re_n$.
\end{proposition}

In some aspects the abelian sandpile measure $\lambda_n,
n\to\infty,$ behaves as a model of statistical mechanics at the
critical point, a phenomenon which is sometimes referred to as
``self-organized criticality" because there is no explicit tuning
of parameters.
 In the physics literature various
critical exponents related to the avalanche behavior are introduced for that model.
One
signature of ``critical behavior" is the presence of power law
decay of correlations for the height $1$ two-point function, as
proven by Majumdar and Dhar in  \cite{md}:

\begin{proposition}\label{power}
There exist constants $c,C>0$ such that
\[
c |x|^{-4}\leq |\lambda_n(\eta(0)=\eta(x)=1) -
\lambda_n(\eta(0)=1)\lambda_n(\eta(x)=1)| \leq C \,|x|^{-4}
\]
for all $x\not= 0$ and $n$ large enough.
\end{proposition}

On the other hand, a contour of 4's completely decouples the
inside and the outside, as we now show. A subset $V\subset \bsZ^2$
is called simply connected if the corresponding
$\hat{V}\subset\R^2$ obtained by ``filling the squares'' of $V$ is
simply connected.

\begin{proposition}\label{decoup}
For $W\subset V_n$ denote by $4_W$ the event that $\eta(x) = 4$ on
$W$. For any simply connected subset $V\subset\bsZ^2$ with
$\overline{V}\subset V_n$:
\be\label{decoupe} \lambda_n
(\eta_V\eta_{V_n\setminus \overline{V}}|4_{\partial V}) =\lambda_n
(\eta_V|4_{\partial V}) \lambda_n (\eta_{V_n\setminus
\overline{V}}|4_{\partial V}) \ee
\end{proposition}
\bpr Denote by $\re^{ext}_{V_n\setminus \overline{V}}$ the set of
configurations which are burnable in $V_n\setminus \overline{V}$
and such that the extension $\eta_{V_n\setminus
\overline{V}}4_{\overline{ V}}$ is burnable in $V_n$. By the
burning algorithm, $\eta_{V}4_{\partial V}\eta_{V_n\setminus
\overline{V}}\in\re_{V_n}$ if and only if $\eta_V\in\re_V$ and
 $\eta_{V_n\setminus \overline{V}}\in\re^{ext}_{V_n\setminus \overline{V}}$.
Therefore
\be
\lambda_n (\eta_V\eta_{V_n\setminus \overline{V}}|4_{\partial V})
=
\frac{ I[\eta_V\in\re_V, \eta_{V_n\setminus \overline{V}}\in
\re^{ext}_{V_n\setminus \overline{V}}]}%
{|\re^{ext}_{V_n\setminus \overline{V}}||\re_V|}
\ee
\be
\lambda_n (\eta_V|4_{\partial V})=\frac{I[\eta_V\in \re_V]}{|\re_V|}
\ee
and
\be
\lambda_n (\eta_{V_n\setminus \overline{V}}|4_{\partial V})
=
\frac{I[\eta_{V_n\setminus \overline{V}}\in \re^{ext}_{V_n\setminus \overline{V}}]|\re_V|}%
{|\re^{ext}_{V_n\setminus \overline{V}}||\re_V|}
\ee
which gives the result.
\epr

\subsection{Sandpile model with Potts interaction}\label{xi}
Define the probability measures $\mu_{\beta,n}^a$ on $\Omega_n$ as
\[
\mu_{\beta,n}^a(\eta) = \frac{\exp(-\beta
H_n (\eta|a)) I[\eta \in
\re_n]}{\Xi_{\beta,n}^a}
\]
where the normalizing constant
$\Xi^a_{\beta,n}$ is
\[
\Xi^a_{\beta,n} \equiv \sum_{\eta \in \re_n} \exp(-\beta
H_n (\eta|a))
\]
Similarly, we define the partition function
$\Xi^a_{\beta,V} $ in an arbitrary finite volume $V$. This
partition function will also be abbreviated as
$\Xi_{a,V}$.\\
$\mu^a_{\beta,n}$ is of course just the original
Potts measure conditioned on being recurrent:
\[
 \mu_{\beta,n}^a(\eta) = \nu_{\beta,n}^a(\eta|\re_n)
\]
 Obviously, at infinite
temperature, $\beta=0$, we recover the stationary measure
$\lambda_n$ of the sandpile model. The constraint $\eta\in\re_n$
can be viewed as introducing an extra nonlocal hard-core
interaction (implicitly given
  by the burning algorithm) but it also breaks the
Potts-symmetry: approximately for $n\uparrow +\infty$,
$\lambda_n(\eta(0)=4)=0.4,
 \lambda_n(\eta(0)=3)=0.3,
  \lambda_n(\eta(0)=2)=0.2,
   \lambda_n(\eta(0)=1)=0.1$ as computed by Priezzhev, \cite{Pri}.

\section{Results}
With boundary condition $a=4$, at
low temperature, the typical configurations of the Potts model on sand look
like an ocean of $4$'s with exponentially damped burnable islands.

\begin{theorem}\label{thm1}
For any $\epsi >0$ there exists $\beta_0\in (0,\infty)$ such that
for all $\beta>\beta_0$ and all $n\in\N$ \be\label{4phase}
\mu^4_{\beta,n} (\eta(0)=4 ) > 1-\epsi \ee Moreover there exists
$c>0$ such that for $\beta>\beta_0$ and $n$ big enough we have the
bound \be\label{expde}
|\mu^4_{\beta,n}(\eta(x)\eta(0))-\mu^4_{\beta,n}(\eta(x))\mu^4_{\beta,n}(\eta(0))|
\leq e^{-c|x|} \ee exponentially small in the distance $|x|$ from
the origin.
\end{theorem}

\eqref{expde} must be contrasted with the situation for $\beta=0$
where there are long range correlations, see Proposition \ref{power}.

Besides ``all 4'', the ``all 3'' is the only other groundstate.  But that
one is unstable:

\bt\label{inst3thm}
 For every $\alpha>0$ there exists
$\beta(\alpha)\in (0,\infty) $ and $c=c(\alpha,\beta)>0$ such that
for all $\beta >\beta(\alpha)$
 \be\label{3thm} \mu^3_{\beta,n} ( |\{ x\in V_n:
\eta(x) =3\}| > \alpha |V_n|) \leq e^{-c|V_n|}
 \ee
for large $n$.
 \et

The next section gives the proof of Theorem \ref{thm1} and
introduces a random cluster representation of the Potts model on
sand. Section \ref{clu} is devoted to the proof of Theorem
\ref{inst3thm}.  It will be seen that, as an extension of Theorem
\ref{inst3thm}, for no matter what boundary conditions, the
density of 4 tends to one with $\beta\uparrow +\infty$.

\section{Random cluster representation}

The volume $\overline{V_n}=V_n\cup\partial V_n$ can be
 considered as a finite graph with the sites
$x\in V_n\cup \partial V_n$ as vertices and with edge set $B_n =
B$ consisting of the nearest neighbor bonds $x\sim y$ where at
least one neighbor is in $V_n$.  We define the {\it sand-Potts}
random cluster measure $\varphi_{p,n}^a=\varphi_p^a$ on this graph
with parameter $p \in [0,1]$ as the probability measure on
$\{0,1\}^B$ which to each $\sigma \in \{0,1\}^B$ assigns
probability
\begin{equation}\label{rcm}
\varphi_p^a(\sigma) = \frac 1{N_p^a} \;[\prod_{e\in B}
p^{\sigma(e)} (1-p)^{1-\sigma(e)}]\, \sum_{\eta\in \re_n} I[\eta
\mbox{ is constant on clusters of}\ \si]
\end{equation}
By cluster we mean a (nearest neighbor) connected component of
sites (including isolated sites) as obtained from the bond
configuration $\sigma$. Bonds for which $\sigma(e)=1$, $\si (e)=0$
are called open, respectively, closed. In this definition, we
assume that the boundary sites are all connected (wired).  All
sites that are connected to the boundary are in the same cluster.
The restriction that $\eta$ is constant on clusters also implies
that $\eta$ is constant equal to $a$ on the cluster of the
boundary. Remember however that $\eta\equiv 2$ and $\eta\equiv 1$
are not in $\re_n$.

\subsection{Stochastic domination}

\begin{lemma}\label{key}
Let $a=3$ or $a=4$.  For every edge $e=\langle xy \rangle$ in $B$
and every $\sigma_{B\setminus e}\in \{0,1\}^{B\setminus \{e\}}$,
\begin{equation}\label{eq}
\varphi_p^a(\sigma(e)=1|\sigma_{B\setminus e}) = p
\end{equation}
if $x$ and $y$ are connected via open edges in $\sigma_{B\setminus e}$. If, on
the other hand,   $x$ and $y$ are not connected via open edges in
$\sigma_{B\setminus e}$,
 then we still have
\begin{equation}\label{upperb}
p\geq \varphi_p^a(\sigma(e)=1|\sigma_{B\setminus e})
\end{equation}
and for  $a=4$,
\begin{equation}\label{lowerb}
\varphi_p^4(\sigma(e)=1|\sigma_{B\setminus e}) \geq \frac{
p}{7-6p}
\end{equation}
\end{lemma}
\bpr
Let $\sigma\in \{0,1\}^{B}$. We write
\begin{equation}\label{notk}
 \sum_{\eta\in \re_n} I[\eta \mbox{ is constant on
clusters}] = k(n,a;\sigma)
\end{equation}
for the number of recurrent configurations that are constant on
the $\sigma-$clusters and fixed equal to $a$ for each site that is
$\sigma-$connected to $\partial V_n$.  It equals $|R_n|$ when all
edges in $\sigma$ are closed.  Obviously, $k(n,a;\sigma)\leq
|R_n|$ and $k(n,a;\sigma)$ is decreasing in $\sigma$ and is
increasing in $a$. For $a=3$ or $4$, $k(n,a;\sigma)\geq 1$.
Continuing with either $a=3$ or $a=4$, we have \be
\frac{\vi^a_p(1_e\si_{B\setminus e})}{\vi^a_p(0_e\si_{B\setminus
e})} = \frac{p \,k(n,a;1_e\si_{B\setminus
e})}{(1-p)\,k(n,a,0_e\si_{B\setminus e})} \ee and hence

\begin{equation}\label{condprob}
\varphi_p^a(\sigma(e)=1|\sigma_{B\setminus e}) =
\frac{1}{1+\frac{\vi^a_p(0_e\si_{B\setminus
e})}{\vi^a_p(1_e\si_{B\setminus e})}} = \frac{p}{p +
(1-p)\frac{k(n,a;0_e\si_{B\setminus
e})}{k(n,a;1_{e}\si_{B\setminus e})}}
\end{equation}
Abbreviate $\sigma^{0,e}=0_e\si_{B\setminus e}$ and
$\sigma^{1,e}=1_{e}\si_{B\setminus e}$; they are both equal to
$\sigma_{B\setminus e}$ off $e$ and $\sigma^{0,e}(e)=0$ and
$\sigma^{1,e}(e)=1$.
\\
To prove the first statement \eqref{eq}: suppose $x,y$ are connected via open
edges in $\sigma_{B\setminus e}$, then every configuration
$\eta\in\re_n$ compatible with $\si$ has $\eta(x)=\eta(y)$, and hence
$k(n,a;\sigma)$
does not depend on $\sigma(e)$ in that case.  However, if $x$ and $y$
are not connected via open edges in $\sigma_{B\setminus e}$, then we must
investigate the effect of merging two clusters.  By making $e$
open, we connect two clusters and we must estimate the new number
of recurrent configurations that are constant on clusters in terms
of the old. Since always
\begin{equation}\label{monot}
k(n,a;\sigma^{0,e}) \geq k(n,a;\sigma^{1,e})
\end{equation}
we obtain \eqref{upperb} from \eqref{condprob}. For the last
statement \eqref{lowerb}, we combine Proposition \ref{domi} with
\eqref{condprob}. Suppose that $\eta\in \re_n$ and is constant on
clusters $C_1$ and $C_2$ taking there the values $a_1$ and $a_2$
respectively. The new configuration $\xi$ defined as
\[
\xi(x) =\eta(x), x\notin C_1\cup C_2, \;\; \xi(x) =
\max\{a_1,a_2\}, x\in  C_1\cup C_2
\]
is still recurrent and is constant on  $C_1\cup C_2$.  Moreover,
if say $C_1$ is the boundary cluster, then necessarily $a_1=4$ and
hence also $\max\{a_1,a_2\}=4$ remains compatible with the
boundary. (This does not work with the boundary condition $a=3$.)
Simple counting shows that the map $\eta \rightarrow \xi$ is at
most seven to one, or
\be\label{bla}
k(n,4;\sigma^{0,e}) \leq 7k(n,4,\sigma^{1,e})
\ee
Combination of (\ref{bla}) and (\ref{condprob}) gives
(\ref{lowerb}).
\epr

 Let $\psi_q$ be the Bernoulli product measure on $\{0,1\}^B$ with
density $q=\psi_q (\si_e=1)$.

\bp \label{ber} The random cluster measure $\varphi_p^4$
stochastically dominates $\psi_q$ with $q= p/(7- 6p)$, i.e.,
$\varphi_p^4(\sigma(e)=1, e\in E) \geq \psi_q(\sigma(e)=1, e\in
E)$ for all edge sets $E$. For $a=3,4,\, \varphi_p^a$ is
stochastically dominated by $\psi_p$. Finally, $\varphi_p^a$
always stochastically dominates $\varphi_{p'}^a$ for $0\leq p'\leq
p\leq 1$. \ep \bpr The first and second statement follows directly
from Lemma \ref{key}, see e.g. Theorem 4.8 in \cite{ghm}. The last
statement follows from \eqref{condprob} and \eqref{monot}. \epr

\subsection{Coupling}

The previous construction and arguments are analogous to and
inspired by the Fortuin-Kasteleyn representation of the standard
Potts model. To recover the $q-$state Potts model, one should
simply replace $\re_n$ in \eqref{notk} with $\Omega_n$. Our next
step, making a coupling between the $\eta-$ and the
$\sigma-$field, is the analogue of the Swendsen-Wang-Edwards-Sokal
coupling, \cite{sw,es}. For a general reference, see \cite{ghm}.

We make a coupling $\pee_{p,n}^a=\pee_p^a$ between the Potts model
on sand and the sand-Potts random cluster measure.  Let $\pee_p^a$
be the probability measure on $\Omega_n\times \{0,1\}^B$
constructed as follows.  Assign first to each site in $V_n$ a
sandvalue according to the probability measure $\lambda_n$ and
each site at the boundary $\partial V_n$ gets the value $a$.
Independently, let each edge in $B$ take the value 0 or 1 with
probabilities $1-p$ and $p$ respectively. Secondly, condition on
the event that no two neighboring sites (including sites at the
boundary) with different heights have an open edge connecting
them.  In a formula,
\[
\pee_p^a(\eta,\sigma) = \frac 1{M_p^a} I[\eta\in \re_n]\; \prod_{e=\langle xy \rangle\in
B}\big[p^{\sigma(e)}
(1-p)^{1-\sigma(e)}\,I[(\eta_x-\eta_y)\sigma(e)=0]\big]
\]
where in the last indicator function it is understood that $\eta(z)=a$ for
$z\in \partial V_n$.

\begin{proposition}\label{coup}
Suppose $\beta = -\ln(1-p)$.  Then, $\mu_{\beta,n}^a$ is the
marginal of $\pee_{p}^a$ projected on $\Omega_n$ and $\varphi_p^a$
is the marginal of $\pee_{p}^a$ projected on $\{0,1\}^B$.
\end{proposition}

\begin{proof}
The proof is by direct computation.  For example,
if we sum over the $\sigma$ we have
\[
\sum_\sigma\prod_{e\in B}\big[p^{\sigma(e)}
(1-p)^{1-\sigma(e)}\,I[(\eta(x)-\eta(y))\sigma(e)=0]\big]=(1-p)^{I[\eta(x)\neq\eta(y)]}
\]
which determines $1-p = \exp[-\beta]$.
\end{proof}

\subsection{Proof of Theorem \ref{thm1}: stability of the 4-phase}
We apply Proposition \ref{coup}:
\begin{eqnarray}\nonumber
&& \mu_{\beta,n}^4(\eta(0)=4)
\nonumber\\
&\geq& \pee_p^4(\eta(0)=4|\ 0 \mbox{ is in the cluster of the
boundary}) \times\nonumber \\\quad&&\varphi_p^4(0 \mbox{ is in the
cluster of the boundary})
\nonumber\\
&= & \varphi_p^4(0 \ \mbox{is in the cluster of the boundary})
\end{eqnarray}
and now Proposition \ref{ber} to conclude that
\[
\mu_{\beta,n}^4(\eta(0)=4) \geq \psi_{p/(7-6p)}(0 \mbox{ is
connected to the boundary})
\]
which goes to one, uniformly in $n$ as $\beta=-\ln(1-p)$ goes to
$+\infty$.  That shows \eqref{4phase}.  For the exponential decay of
correlations, \eqref{expde}, observe that by the very
 same argument as above one shows that for large $\beta$, in
$\mu_{\beta,n}^4$ there is percolation of 4's uniformly in $n$.
 For each $n$ the
$\mu_{\beta,n}^4-$probability that there is a nearest-neighbor
path of $4$'s connecting the origin with the boundary $\partial
V_n$ is not smaller than the percolation probability in the
Bernoulli bond process with occupation probability $(1-
\exp[-\beta])/(1 + 6\exp[-\beta])$. Moreover, we can always
consider a rectangle parallel to $V_n$ between the origin and site
$x$ with one side proportional to $|x|$ and the other side equal
to $n$. Again by the same domination argument, the
$\mu_{\beta,n}^4-$probability that in that rectangle, there is
percolation of 4's from one side of $V_n$ to the opposite side is
not smaller than $1 - \exp[-c|x|]$ with $c\uparrow+\infty$ as
$\beta\uparrow +\infty$, uniformly in $n$.  Therefore, denoting
that event by ``crossing', we have that
\[
|\mu^4_{\beta,n}(\eta(x)\eta(0)) -
\mu^4_{\beta,n}(\eta(x)\eta(0)|\mbox{crossing})| \leq \exp[-c|x|]
\]
That can be combined with Proposition \ref{decoup}; if the origin
and the site $x$ are separated by a path of 4's, then they are
independent:
\[
\mu^4_{\beta,n}(\eta(x)\eta(0)|\mbox{crossing}) =
\mu^4_{\beta,n}(\eta(x)|\mbox{crossing})
\mu^4_{\beta,n}(\eta(0)|\mbox{crossing})
\]
For each of the two factors we can use the previous argument to
conclude the proof.

\section{Instability of the 3-phase}\label{clu}
In this section we prove Theorem \ref{inst3thm}.

 The main idea is to consider the restricted ensemble defined
below (following ideas from \cite{ds}). Let $V$ be a finite
(large) volume and $\eta\in\Omega_V$ a configuration in the volume
$V$. Define the contours of $\eta$ as connected components of
edges separating the different values of $\eta$ in neighboring
sites (we now consider these sites as centers of unit squares and
the edges are the sides of these squares). We suppose that the
configuration $\eta\in\re_V$ is burnable. The set of
configurations  that have contours with lengths not exceeding 12
not touching $\partial V$ is denoted by $\re^{12}_V$. That set of
configurations defines a restricted ensemble. It is easy to see
that we have the following alternative: either the configuration
$\eta$ is a sea of $4$ with islands made from contours with length
$\leq 12$, or it is a sea of $3$ with such islands: \be \re^{12}_V
= \re^{12}_{V,4}\cup \re^{12}_{V,3} \ee Notice that a sea of $2$
or $1$ is not possible by burnability.

We list all subconfigurations of energy $\leq 12$ in the sea of 4
and in the sea of 3. Of course, these islands have to be burnable.
In the figure, empty cells
are part of the sea.\\
\subsection{Islands in the sea of 4}
\ben
\item
Energy =4:\\
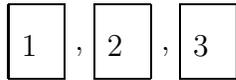
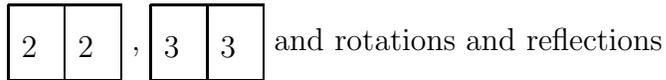
\begin{tabular}{|p{0.4cm}|}
\hline 1\\
\hline
\end{tabular}
,
\begin{tabular}{|p{0.4cm}|}
\hline 2\\
\hline
\end{tabular}
,
\begin{tabular}{|p{0.4cm}|}
\hline 3\\
\hline
\end{tabular}
\item
Energy =6:\\
\begin{tabular}{|p{0.4cm}|p{0.4cm}|}
\hline 2 & 2\\
\hline
\end{tabular}
,
\begin{tabular}{|p{0.4cm}|p{0.4cm}|}
\hline 3 & 3\\
\hline
\end{tabular}
and rotations and reflections\\
(this remark
will be omitted from here on)
\item Energy = 7\\
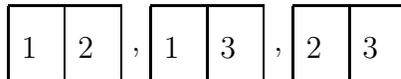
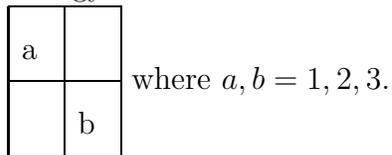
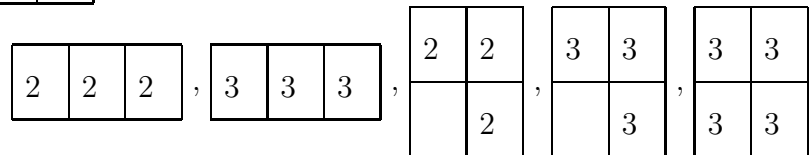
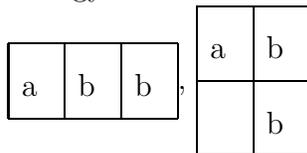
\begin{tabular}{|p{0.4cm}|p{0.4cm}|}
\hline 1 & 2\\
\hline
\end{tabular}
,
\begin{tabular}{|p{0.4cm}|p{0.4cm}|}
\hline 1 & 3\\
\hline
\end{tabular}
,
\begin{tabular}{|p{0.4cm}|p{0.4cm}|}
\hline 2 & 3\\
\hline
\end{tabular}
\item Energy =8\\
\begin{tabular}{|p{0.4cm}|p{0.4cm}|}
\hline a & \ \\
\hline \ & b \\
\hline
\end{tabular}
where $a,b=1,2,3$.

\begin{tabular}{|p{0.4cm}|p{0.4cm}|p{0.4cm}|}
\hline 2 & 2 &2\\
\hline
\end{tabular}
,
\begin{tabular}{|p{0.4cm}|p{0.4cm}|p{0.4cm}|}
\hline 3 & 3 &3\\
\hline
\end{tabular}
,
\begin{tabular}{|p{0.4cm}|p{0.4cm}|}
\hline 2 &2 \ \\
\hline \ & 2 \\
\hline
\end{tabular}
,
\begin{tabular}{|p{0.4cm}|p{0.4cm}|}
\hline 3 & 3\ \\
\hline \ & 3 \\
\hline
\end{tabular}
,
\begin{tabular}{|p{0.4cm}|p{0.4cm}|}
\hline 3 & 3 \\
\hline 3 & 3 \\
\hline
\end{tabular}

\item Energy =9\\
\begin{tabular}{|p{0.4cm}|p{0.4cm}|p{0.4cm}|}
\hline a & b &b\\
\hline
\end{tabular},
\begin{tabular}{|p{0.4cm}|p{0.4cm}|}
\hline a & b\ \\
\hline \ & b \\
\hline
\end{tabular}

where $a=1,2,3$, $b=2,3$, $a\not= b$.
\item
Energy =10\\
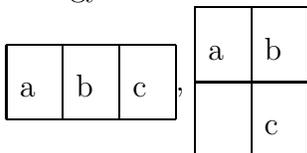
\begin{tabular}{|p{0.4cm}|p{0.4cm}|p{0.4cm}|}
\hline a & b &c\\
\hline
\end{tabular},
\begin{tabular}{|p{0.4cm}|p{0.4cm}|}
\hline a & b \\
\hline \ & c\\
\hline
\end{tabular}

where $a\not= b$, $b\not= c$; $a,b,c= 1,2,3$ and if $a=c=1$ then
$b=3$.

\begin{tabular}{|p{0.4cm}|p{0.4cm}|p{0.4cm}|}
\hline a & a & \ \\
\hline \ & \ & b \\
\hline
\end{tabular}
where $a=2,3$, $b=1,2,3$.

\begin{tabular}{|p{0.4cm}|p{0.4cm}|}
\hline a & a \\
\hline \ & a\\
\hline \ & a\\
\hline
\end{tabular},
\begin{tabular}{|p{0.4cm}|p{0.4cm}|}
\hline \ & a \\
\hline a & a\\
\hline \ & a\\
\hline
\end{tabular},
\begin{tabular}{|p{0.4cm}|p{0.4cm}|p{0.4cm}|p{0.4cm}|}
\hline a & a & a & a\\
\hline
\end{tabular},
\begin{tabular}{|p{0.4cm}|p{0.4cm}|}
\hline a & \ \\
\hline a & a\\
\hline \ & a\\
\hline
\end{tabular},

where $a=2,3$.

\begin{tabular}{|p{0.4cm}|p{0.4cm}|}
\hline 2 & 2 \\
\hline 3 & 2\\
\hline
\end{tabular},
\begin{tabular}{|p{0.4cm}|p{0.4cm}|}
\hline 3 &3 \\
\hline 2 & 3\\
\hline
\end{tabular},
\begin{tabular}{|p{0.4cm}|p{0.4cm}|}
\hline 3 & 3 \\
\hline 1 & 3\\
\hline
\end{tabular},
\begin{tabular}{|p{0.4cm}|p{0.4cm}|p{0.4cm}|}
\hline \ & 3 & 3 \\
\hline 3 & 3 & 3\\
\hline
\end{tabular},
\begin{tabular}{|p{0.4cm}|p{0.4cm}|p{0.4cm}|}
\hline 3 & 3 & 3 \\
\hline 3 & 3 & 3\\
\hline
\end{tabular}.

\item Energy =11

\begin{tabular}{|p{0.4cm}|p{0.4cm}|p{0.4cm}|}
\hline a & b &\ \\
\hline \ & \ & c\\
\hline
\end{tabular}
where $a,b,c=1,2,3$, $a\not= b$.

\begin{tabular}{|p{0.4cm}|p{0.4cm}|}
\hline b & b \\
\hline a & c\\
\hline
\end{tabular}
where $b=2,3$, $a,c=1,2,3$, $a\not= b$, $a\not= c$, $b\not= c$.

\begin{tabular}{|p{0.4cm}|p{0.4cm}|p{0.4cm}|}
\hline a & a & a\\
\hline \ & \ & b\\
\hline
\end{tabular},
\begin{tabular}{|p{0.4cm}|p{0.4cm}|p{0.4cm}|}
\hline a & a & a\\
\hline \ & b &\ \\
\hline
\end{tabular}

\begin{tabular}{|p{0.4cm}|p{0.4cm}|}
\hline a & a \\
\hline \ & a\\
\hline \ & b\\
\hline
\end{tabular},
\begin{tabular}{|p{0.4cm}|p{0.4cm}|p{0.4cm}|}
\hline a & a & b\\
\hline \ & a & \ \\
\hline
\end{tabular}

\begin{tabular}{|p{0.4cm}|p{0.4cm}|p{0.4cm}|}
\hline a & a & \ \\
\hline \ & a & b \\
\hline
\end{tabular},
\begin{tabular}{|p{0.4cm}|p{0.4cm}|p{0.4cm}|p{0.4cm}|}
\hline a & a & a & b\\
\hline
\end{tabular}

where $a=2,3$, $b=1,2,3$, $b\not= a$.

\begin{tabular}{|p{0.4cm}|p{0.4cm}|p{0.4cm}|}
\hline 3 & 3 & \ \\
\hline 3 & 3 & a \\
\hline
\end{tabular},
where a=1,2.

\item Energy =12

\begin{tabular}{|p{0.4cm}|p{0.4cm}|p{0.4cm}|}
\hline a & \ & \ \\
\hline \ & b & \ \\
\hline \ & \ & c \\
\hline
\end{tabular},
\begin{tabular}{|p{0.4cm}|p{0.4cm}|p{0.4cm}|}
\hline a & \ & c\\
\hline \ & b & \ \\
\hline
\end{tabular},
where $a,b,c= 1,2,3$.

\begin{tabular}{|p{0.4cm}|p{0.4cm}|}
\hline a & b \\
\hline c & a \\
\hline
\end{tabular}

where $a,b,c = 1,2,3$, $a\not= b$, $a\not= c$, if $a=1$ then
$b=c=3$, if $a=2$, then $b=3$ or $c=3$.

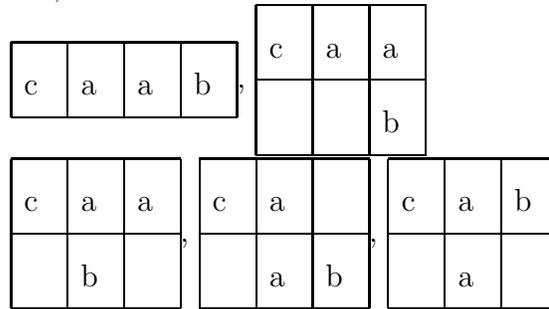
\begin{tabular}{|p{0.4cm}|p{0.4cm}|p{0.4cm}|p{0.4cm}|}
\hline c & a & a &b\\
\hline
\end{tabular},
\begin{tabular}{|p{0.4cm}|p{0.4cm}|p{0.4cm}|}
\hline c & a & a\\
\hline \ & \ & b \\
\hline
\end{tabular}

\begin{tabular}{|p{0.4cm}|p{0.4cm}|p{0.4cm}|}
\hline c & a & a\\
\hline \ & b & \ \\
\hline
\end{tabular},
\begin{tabular}{|p{0.4cm}|p{0.4cm}|p{0.4cm}|}
\hline c & a & \ \\
\hline \ & a & b \\
\hline
\end{tabular},
\begin{tabular}{|p{0.4cm}|p{0.4cm}|p{0.4cm}|}
\hline c & a & b\\
\hline \ & a & \ \\
\hline
\end{tabular}

where a=2,3, $b,c = 1,2,3$, $a\not= b$, $a\not= c$; if $a=2$, then
$b=3$ or $c=3$.

\begin{tabular}{|p{0.4cm}|p{0.4cm}|p{0.4cm}|}
\hline 3 & 3 & \ \\
\hline 3 & 3 & \ \\
\hline \ & \ & a \\
\hline
\end{tabular}
for $a=1,2,3$.

\begin{tabular}{|p{0.4cm}|p{0.4cm}|p{0.4cm}|p{0.4cm}|}
\hline a & a & b & c\\
\hline
\end{tabular},
\begin{tabular}{|p{0.4cm}|p{0.4cm}|p{0.4cm}|}
\hline a & a & b \\
\hline \ & \ & c \\
\hline
\end{tabular},
\begin{tabular}{|p{0.4cm}|p{0.4cm}|p{0.4cm}|}
\hline a & a & \ \\
\hline \ & b & c \\
\hline
\end{tabular},
\begin{tabular}{|p{0.4cm}|p{0.4cm}|}
\hline a & a  \\
\hline \ & b  \\
\hline \ & c \\
\hline
\end{tabular}

where $a,b,c= 1,2,3$, $a\not= 1$, $a\not= b$, $b\not= c$.

\begin{tabular}{|p{0.4cm}|p{0.4cm}|p{0.4cm}|}
\hline \ & \ & b\\
\hline a & a & \ \\
\hline \ & a & \ \\
\hline
\end{tabular},
\begin{tabular}{|p{0.4cm}|p{0.4cm}|p{0.4cm}|}
\hline a & a & \ \\
\hline \ & a & \ \\
\hline \ & \ & b \\
\hline
\end{tabular}

\begin{tabular}{|p{0.4cm}|p{0.4cm}|}
\hline a & a  \\
\hline \ & a  \\
\hline b & \  \\
\hline
\end{tabular}

where $a=2,3$ and $b=1,2,3$.

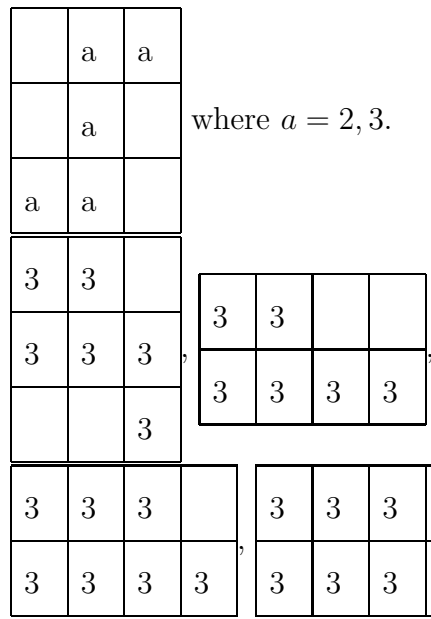
\begin{tabular}{|p{0.4cm}|p{0.4cm}|}
\hline a & a  \\
\hline \ & a \\
\hline a &  a \\
\hline
\end{tabular},
\begin{tabular}{|p{0.4cm}|p{0.4cm}|p{0.4cm}|p{0.4cm}|p{0.4cm}|}
\hline a & a & a & a& a  \\
\hline
\end{tabular}

\begin{tabular}{|p{0.4cm}|p{0.4cm}|p{0.4cm}|}
\hline a & a & a \\
\hline \ & \  & a\\
\hline \ &  \ & a \\
\hline
\end{tabular},
\begin{tabular}{|p{0.4cm}|p{0.4cm}|}
\hline a & a  \\
\hline \ & a \\
\hline \ &  a \\
\hline \ & a\\
\hline
\end{tabular}

\begin{tabular}{|p{0.4cm}|p{0.4cm}|p{0.4cm}|}
\hline \ & a & a  \\
\hline \ & a & \  \\
\hline a &  a & \ \\
\hline
\end{tabular}
where $a=2,3$.

\begin{tabular}{|p{0.4cm}|p{0.4cm}|p{0.4cm}|}
\hline 3 & 3 & \   \\
\hline 3 & 3 & 3 \\
\hline \  &  \ & 3 \\
\hline
\end{tabular},
\begin{tabular}{|p{0.4cm}|p{0.4cm}|p{0.4cm}|p{0.4cm}|}
\hline 3 & 3 & \ & \   \\
\hline 3 & 3 & 3& 3\\
\hline
\end{tabular},

\begin{tabular}{|p{0.4cm}|p{0.4cm}|p{0.4cm}|p{0.4cm}|}
\hline 3 & 3 & 3 & \   \\
\hline 3 & 3 & 3 & 3 \\
\hline
\end{tabular},
\begin{tabular}{|p{0.4cm}|p{0.4cm}|p{0.4cm}|p{0.4cm}|}
\hline 3 & 3 & 3 & 3   \\
\hline 3 & 3 & 3& 3\\
\hline
\end{tabular}.

\een
 \subsection{Islands in the sea of 3}
 \bl \label{l-1} Islands
with energy $\leq 11$ in the sea of 3 are in one-to-one
correspondence with islands with the same energy of 4. The
correspondence is given by substitution $3\longleftrightarrow 4$.
\el
 \bl \label{l0}The same
correspondence can be applied to all islands with energy =12, with
two exceptions:

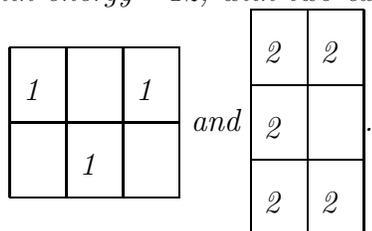
\begin{tabular}{|p{0.4cm}|p{0.4cm}|p{0.4cm}|}
\hline 1 & \ & 1   \\
\hline \ & 1 & \ \\
\hline
\end{tabular}
and
\begin{tabular}{|p{0.4cm}|p{0.4cm}|}
\hline 2 & 2    \\
\hline 2 & \  \\
\hline 2  & 2  \\
\hline
\end{tabular}.

The correspondence gives all islands in the sea of $3$ with energy
12. \el
 \bpr
  This follows from the fact that

\begin{tabular}{|p{0.4cm}|p{0.4cm}|p{0.4cm}|}
\hline 1 & 4 & 1   \\
\hline \ & 1 & \ \\
\hline
\end{tabular}
and
\begin{tabular}{|p{0.4cm}|p{0.4cm}|}
\hline 2 & 2\\
\hline 2 & 4 \\
\hline 2 & 2\\
\hline
\end{tabular}

are burnable but

\begin{tabular}{|p{0.4cm}|p{0.4cm}|p{0.4cm}|}
\hline 1 & 3& 1   \\
\hline \ & 1 & \ \\
\hline
\end{tabular}
and
\begin{tabular}{|p{0.4cm}|p{0.4cm}|}
\hline 2 & 2\\
\hline 2 & 3 \\
\hline 2 & 2\\
\hline
\end{tabular}

are not burnable. \epr

\subsection{Free energy estimates}
Define $\Xi^{(12)}_{4,V}$ and $\Xi^{(12)}_{3,V}$ to be the
partition functions of the ensembles $\re^{12}_{4,V}$,
respectively $\re^{12}_{3,V}$. \bl \label{l1} There exists
$\beta_0$  such that for any $\beta> \beta_0$ there exists a
constant $c>0$ such that for any sufficiently large volume $V$
\be\label{difpart} \log \Xi^{(12)}_{4,V} - \log \Xi^{(12)}_{3,V}
\geq c|V| e^{-12\beta} \ee \el
 \bpr
  Define the extended ensemble
$\re^{12,ext}_{V,3}$, consisting of {\em all} configurations of
disjoint burnable islands with energy $\leq 12$ in the sea of $3$
in volume $V$. Evidently
\[
\re^{12}_{V,3}\subset \re^{12,ext}_{V,3}
\]
and the inclusion is strict because e.g.

\begin{tabular}{|p{0.4cm}|p{0.4cm}|p{0.4cm}|p{0.4cm}|}
\hline  2& 3 & 3 & 2 \\
\hline  2 & 3 & 3 & 2  \\
\hline
\end{tabular}

is a subconfiguration allowed in the ensemble
$ \re^{12,ext}_{V,3}$  but not in the ensemble
$\re^{12}_{V,3}$. Indeed, it is not burnable but the
island

\begin{tabular}{|p{0.4cm}|}
\hline  2\\
\hline  2\\
\hline
\end{tabular} is burnable.

If we analogously define $\re^{12,ext}_{V,4}$, then in fact
$\re^{12,ext}_{V,4}=\re^{12}_{V,4}$ because a configuration of
(disjoint) islands in a sea of $4$ is burnable if and only if all
the individual islands are burnable in the sea of 4, see
Proposition \ref{decoup}.

Therefore
\[
\Xi^{(12,ext)}_{3,V} \geq \Xi^{(12)}_{3,V}
\]
and we can obtain Lemma \ref{l1} from the following
\bp\label{clustp}
 \be\label{clust}
 \log\Xi^{(12)}_{4,V}-\log \Xi^{(12,ext)}_{3,V} \geq c|V|
e^{-12\beta} \ee \ep The proof is an application of the usual
polymer expansion (see \cite{Kot,Sei}) because islands in the sea
of 4 do not interact (again from Proposition \ref{decoup}).
Comparing the expansions of $\log\Xi^{(12)}_{4,V}$ and $\log
\Xi^{(12,ext)}_{3,V}$ term by term, and using Lemma \ref{l-1} and
Lemma \ref{l0}, we obtain a difference in the terms of order
$e^{-12\beta}$ and no difference in previous terms. In the same
way it is easy to prove a weaker inequality valid for all finite
$V\subset\bsZ^d$ \bp\label{clustpp} For $\beta>\beta_0$ there
exist $c>0$, $f\in\R$ such that for any finite $V\subset\bsZ^d$:
\be\label{clustppp} \log\Xi^{(12)}_{4,V}-\log \Xi^{(12)}_{3,V}\geq
\log\Xi^{(12)}_{4,V}-\log \Xi^{(12,ext)}_{3,V} \geq c|V|
e^{-12\beta} - f|\partial V| \ee \ep \epr

\subsection{Proof of Theorem \ref{inst3thm}}

Again, the connected components of the set of lattice edges
separating different values of neighboring spins are called
contours.  Each contour $\gamma$ will be associated with values of
spins on all squares touching $\gamma$. Consider the set $\Gamma =
\{ \gamma_1,\ldots,\gamma_n\}$ of all those contours of a given
configuration $\eta\in\re_{V,3}$ which have energy (length)
exceeding 12, $|\gamma_i|>12$. We call $\Gamma$ the polycontour of
$\eta$. The polycontour separates the volume $V$ into a finite
number of connected domains $V_1,\ldots,V_m$. To each domain $V_i$
is associated the value of the spin $\kappa_i$, induced by
contours $\gamma_j$ neighboring this domain (so $\kappa_i= \kappa$
if the spins in the inner boundary of $V_i$ are equal to
$\kappa$). If $\Gamma=\emptyset$ then there is the domain $V$, to
which is associated $\kappa= 3$. Denote the band consisting of
squares touching $\Gamma$ by $[\Gamma]$. From burnability it
follows that all domains $V_i$ for which $\kappa_i= 1,2$ are
contained in $[\Gamma]$. We denote by $V^{(3)}$ the union of those
$V_i$ for which $\kappa_i=3$, and by $V^{(4)}$ the union of those
$V_i$ for which $\kappa_i=4$. Now $\zeta$, the number of $3$'s in
$V$, equals
\[
\zeta= \zeta^{(3)}+ \zeta^{(4)}
\]
where $\zeta^{(3)}$, respectively $\zeta^{(4)}$ denotes the number
of 3's in $V^{(3)}$, respectively $V^{(4)}$. Therefore
\[
\mbox{Prob}[\zeta \geq\alpha|V|] \leq
\mbox{Prob}[\zeta^{(3)}\geq\frac{\alpha}{2} |V|] +
\mbox{Prob}[\zeta^{(4)}\geq\frac{\alpha}{2} |V|]
\]
We will separately estimate the two terms in the right-hand side.
The last one is the easiest. By definition of $V^{(4)}$, all spin
values equal to 3 in $V^{(4)}$ are contained in the islands listed
above. Each island contains $\leq 8$ squares and has an energy
$\geq 4$. So the energy per square is $\geq 1/2$. If
$\zeta^{(4)}\geq|V|\alpha/2$, then the energy of the configuration
is $\geq \alpha |V|/4$. Note that $\Xi_{3,V}\geq 1$ because the
configuration $\equiv 3$ has zero energy. So the probability of
the configuration $\eta$ is less than or equal $\exp(-\beta \alpha
|V|/4)$. As the total number of configurations does not exceed
$4^{|V|}$, \be
 \mu_{\beta,V}^3 \left(\zeta^{(4)}\geq
\frac{\alpha}2 |V|\right) \leq 4^{|V|} \exp\left(-\beta
\frac{\alpha}{4} |V|\right) \ee
 which, when $\alpha\beta > 4\log
4$, tends to zero as $|V|\to\infty$.

The rest of the proof is an estimate of $\mu_{\beta,V}^3
(\zeta^{(3)}\geq\alpha |V|/2)$.

Note that $\zeta^{(3)} \leq |V^{(3)}|$, and hence
 \be
\mu_{V,\beta}^3 (\zeta^{(3)}\geq\frac{\alpha}2 |V|) \leq
\mu_{V,\beta}^3( |V^{(3)}|\geq\frac{\alpha}2 |V|) =
\frac{Z_\alpha}{\Xi_{3,V}} \ee where $Z_\alpha$ is the partition
function of those configurations $\eta\in\re_{V,3}$ for which $
|V^{(3)}|\geq\frac{\alpha}2 |V|$.

It is easy to see that \be\label{sumg} Z_\alpha \leq \sum_{\Gamma}
e^{-\beta |\Gamma|}\Xi^{(12)}_{3,V^{(3)}} \Xi^{(12)}_{4,V^{(4)}}
\ee where the sum runs over those polycontours $\Gamma$, for which
$|V^{(3)}|\geq\frac{\alpha}2 |V|$ ($\Gamma=\emptyset$, for which
$V^{(3)} =V$ is included).

Let us now estimate $\Xi_{3,V}$. Consider only those
configurations $\eta\in\re_{V,3}$ which also belong to
$\re^{12}_{V',4}$, where $V'$ is $V$ without squares touching
$\partial V$. Then we have \be \Xi_{3,V} \geq e^{-3\beta|\partial
V|} \Xi^{(12)}_{4,V'} \ee From the cluster expansion (or by direct
counting), \be \Xi^{(12)}_{4,V'}\geq e^{-h |\partial V|}
\Xi^{(12)}_{4,V}
\ee
 for some constant $h>0$. Therefore, \be
\Xi_{3,V} \geq e^{-d |\partial V|} \Xi^{(12)}_{4,V} \ee for some
constant $d$ (depending on $\beta$). So we obtain \be
\frac{Z_\alpha}{\Xi_{3,V}} \leq e^{d |\partial V|} \frac{
\sum_{\Gamma} e^{-\beta |\Gamma|}\Xi^{(12)}_{3,V^{(3)}}
\Xi^{(12)}_{4,V^{(4)}}}{\Xi^{(12)}_{4,V}} \ee where the sum over
$\Gamma$ is as in \eqref{sumg}.  In the right-hand side use that
for every $\xi_1\in \re^{12}_{V^{(3)},4}$ and for every $\xi_2\in
\re^{12}_{V^{(4)},4}$, there exists a $\zeta \in \re^{12}_{V,4}$
that coincides with $\xi_1$ on $V^{(3)}$ and with  $\xi_2$ on
$V^{(4)}$.  Hence,
 \be
 \Xi^{(12)}_{4,V} \geq
\Xi^{(12)}_{4,V^{(3)}}\,\Xi^{(12)}_{4,V^{(4)}} \ee
 and thus
  \be
\frac{Z_\alpha}{\Xi_{3,V}} \leq e^{d |\partial V|}  \sum_{\Gamma}
e^{-\beta
|\Gamma|}\frac{\Xi^{(12)}_{3,V^{(3)}}}{\Xi^{(12)}_{4,V^{(3)}}} \ee
 Now use the
inequality \eqref{clustppp} in order to obtain the following
estimate \be \mbox{(4.6) } \leq  \mbox{(4.13) } \leq e^{d|\partial
V|} \sum_{\Gamma} e^{-\beta |\Gamma|} e^{f|\partial V^{(3)}|}
e^{-c |V^{(3)}| e^{-12\beta}} \ee But evidently $|\partial
V^{(3)}|\leq |\partial V| + |\Gamma|$, and $V^{(3)}\geq\alpha
|V|/2$, so we have the estimate \be e^{(d+f)|\partial V|}
e^{-\frac12 \alpha c |V| e^{-12\beta}}\sum_{\Gamma}
e^{-\beta'|\Gamma|} \ee where $\beta'=\beta-f$. To estimate the
last sum is standard combinatorics:
 \be
  \sum_{\Gamma} e^{-\beta'|\Gamma|} \leq
\prod_{x\in X} \left(1 + \sum_{\gamma\ni x, |\gamma| > 12}
e^{-\beta' |\gamma|}\right)
\ee
 where $X$ is the set of all
vertices --- ends of edges inside $V$, $|X|\leq |V|$. A classical
Peierls' estimate gives \be \sum_{\gamma\ni x, |\gamma| > 12}
e^{-\beta' |\gamma|} \leq H' e^{- 13\beta'} \leq H e^{- 13\beta}
\ee for some constants $H=H'\exp[13 f]$. Since
\[
1+  H e^{- 13\beta}\leq e^{H  e^{- 13\beta}}
\]
we obtain the estimate
\be
 \mbox{(4.6) } \leq
e^{(d+f)|\partial V|} e^{-\frac12 \alpha c |V| e^{-12\beta}} e^{H
|V|  e^{- 13\beta}} \ee If $\beta > \log [2H/\alpha c]$, this
product tends to zero as $|V|\to\infty$, $|\partial V|/|V|\to 0$.
That together with (4.5) concludes the proof.

 \br\nonumber
 To
prove that the density of 1's or 2's goes to zero is even easier:
Let $N$ be the number of 1's and 2's in $V$, $N=N^{(3,4)} +
N^{[\Gamma]}$ where $N^{(3,4)}$ is the number of 1's and 2's in
$V^{(3)}\cup V^{(4)}$ and $N^{[\Gamma]}$ is the number of 1's and
2's in $[\Gamma]$. Observe that $N^{[\Gamma]}\leq 2|\Gamma|$, and
the 1's and 2's in $V^{(3)}\cup V^{(4)}$ are contained in islands.
For the islands, the energy per square is $\geq 1/2$.  The energy
of a configuration is equal to $|\Gamma|$ plus the energy of the
islands.  Hence, if $N\geq \alpha |V|$, then that energy is not
less than $\alpha |V|/2$. We can then do as in (4.5) to prove that
\[ \mu_{\beta,V}^3(N \geq \alpha |V|) \rightarrow 0 \quad\mbox{ as }
|V|\rightarrow +\infty
\]
 whenever $\alpha \beta > 2 \log 4$.\\
 The same
 proof as above remains valid for the ensemble $\re_{V,4}$.
In fact, it is valid modulo straightforward modifications, for
general boundary conditions as well. The proof for the ensemble
with general boundary conditions $\xi$ only needs small
modifications concerning the islands touching the boundary.
  \er


\begin{thebibliography}{99}

\bibitem{btw}
Bak, P., Tang, K. and  Wiesenfeld, K., {\em Self-Organized
Criticality}, Phys. Rev. A  {\bf 38}, 364--374 (1988).


\bibitem{dhar}
Dhar, D., {\em The Abelian Sandpiles and Related Models}, Physica
A {\bf 263}, 4--25 (1999).

\bibitem{ds}
Dinaburg, E.L. and Sinai, Ya.G., {\em An analysis of ANNNI model
by Peierl's contour method}, Commun. Math. Phys. {\bf 98},
119--144 (1985).



\bibitem{md}
Majumdar, S.N. and Dhar, D., Physica A {\bf 185}, 129 (1992), J.
Phys. A: Math. Gen. {\bf 24} L357 (1991).

\bibitem{es}
Edwards, R.G. and Sokal, A.D., {\em Generalization of the
Fortuin-Kasteleyn-Swendsen-Wang representation and Monte Carlo
algorithm} Phys. Rev. D {\bf 38}, 2009--2012 (1988).

\bibitem{ghm}
Georgii, H.-O., H\"aggstr\"om, O., Maes, C., {\em  The random
geometry of equilibrium phases, Phase Transitions and Critical
Phenomena}, Vol. {\bf 18}, Eds C. Domb and J.L. Lebowitz (Academic
Press, London), 1--142 (2001).



\bibitem{IP}
E.V. Ivashkevich, Priezzhev, V.B., {\em Introduction to the
sandpile model}, Physica A {\bf 254}, 97--116 (1998).

\bibitem{Kot} M. Zahradnik, {\em Analyticity of
low-temperature phase diagrams of lattice spin models}, J. Stat.
Phys. {\bf 47}, 725--755 (1987).


\bibitem{meest}
Meester, R., Redig, F. and Znamenski, D., {\em The abelian
sandpile; a mathematical introduction}, Markov Proc. Rel. Fields,
{\bf 7}, 509--523 (2002).

\bibitem{Pri}
Priezzhev, V.B., {\em Structure of Two Dimensional Sandpile. I.
Height Probabilities}, J. Stat. Phys {\bf 93}

\bibitem{Sei}
Seiler, E., Gauge Theories as a Problem of Constructive Quantum
Field Theory and Statistical Mechanics, Lecture Notes in Physics
{\bf 159}. Springer, Berlin, 1982.


\bibitem{Speer}
Speer, E., {\em Asymmetric Abelian Sandpile Models}, J. Stat.
Phys. {\bf 71}, 61--74 (1993).

\bibitem{sw}
Swendsen, R.H. and Wang, J.-S., {\em Nonuniversal critical
dynamics in Monte Carlo simulations}, Phys. Rev. Lett. {\bf 58}.
86--88 (1987).



\end{thebibliography}
\end{document}